\documentclass{article}
\usepackage{amsmath}
\usepackage{times}
\usepackage{amssymb}
\usepackage{graphicx}
\usepackage{cite}
\usepackage[toc,page]{appendix}

\begin{document}
	\title{Instability of a Vortex Ring due to Toroidal Normal Fluid Flow in Superfluid $^4$He}
	\author{Bhimsen K. Shivamoggi \\ University of Central Florida \\ Orlando, Fl 32816-1364}
	\date{}
	\maketitle
	
	\begin{abstract}
		Vortex rings self-propelling in superfluid $^4$He are shown to be driven unstable by a \textit{toroidal} normal fluid flow. This instability has qualitative similarities with the \textit{Donnelly-Glaberson instability} of Kelvin waves on a vortex filament driven by the normal fluid flow \textit{along} the vortex filament. The growth rate of the present instability is found to be independent of the radius of the vortex ring.
	\end{abstract}

	\section{Introduction}
	Following Landau \cite{Landau}, one considers the superfluid $^4$He below the \textit{Lambda} point as an inviscid, irrotational fluid with thermal excitations superposed on that fluid. These excitations are modeled by a normal fluid whose interactions via \textit{mutual friction} with the superfluid are mediated by vortices. The \textit{mutual friction}\footnote{The mutual friction is known (Schwarz \cite{Schwarz2}, Shivamoggi \cite{Shivamoggi3}, \cite{Shivamoggi4}) to play the dual roles of driving force and drag force and hence to produce both growth and decay of the vortex line length. We do not have an adequate understanding of the underlying roton-vortex scattering process in superfluid $^4$He yet (Donnelly \cite{Donnelly}).} models the scattering of thermal excitations by the vortices (Feynman \cite{Feynman}) and was confirmed experimentally via the second sound attenuation in liquid $^4$He (Hall and Vinen \cite{HallVinen7}, \cite{HallVinen8}). Vinen \cite{Vinen9} gave a phenomenological derivation of the mutual friction force.  Vortices in superfluid $^4$He are, as Onsager \cite{Onsager} suggested, linear topological defects with the superfluid density vanishing  at the vortex core and the circulation around a vortex line quantized, which was confirmed experimentally by Vinen \cite{Vinen11}. Thanks to the circulation quantization constraint, as Feynman \cite{Feynman} suggested, the only possible turbulent motion in a superfluid is a disordered motion of tangled vortex lines. 
	
	Vortex rings are self-propelled three-dimensional toroidal structures (Saffman \cite{Saffman})\footnote{In his vortex theory of matter, Kelvin \cite{Kelvin} proposed to explain the behavior of atoms by considering them as vortex rings and knots in ether.}, which are believed to play a key role in the mechanism of superfluid turbulence (Svistunov \cite{Svistunov}, Vinen \cite{Vinen14}). Superfluid turbulence has been suggested to be the kinetics of merging and splitting vortex rings rather than the kinetics of tangled vortex lines (Nemirovski \cite{Nemirovski}, Walmsley and Golov \cite{WalmsleyGolov}). The emission of vortex rings from reconnections between vortex filaments is believed to facilitate the energy transfer from large-scale quasi-classical motion to small-scale \textit{Kelvin-wave} cascade in the ultra-low temperature regime (Svistunov \cite{Svistunov}, Zhu et al. \cite{Zhu}, Kursa et al. \cite{Kursa}). Walmsley et al. \cite{Walmsley} generated superfluid turbulence experimentally via collisions in a beam of unidirectional vortex rings in superfluid $^4$He in the limit of zero temperature ($0.05K$). Rayfield and Reif \cite{RayfieldReif} used ions coming from a radioactive cathode to produce vortex rings and gave direct experimental confirmation for the existence of quantized circulation. The vortex ring nucleation process sets in when the ions are accelerated by imposed electric fields and reach a critical velocity (Walmsley and Golov \cite{WalmsleyGolov})\footnote{Low temperature conditions favor the vortex-ring generation by keeping the thermal excitations (phonons) sufficiently low and hence the energy loss small. At higher temperatures, rotons appear and cause large energy losses.}. The motion of the vortex rings was controlled and detected by tagging each ring with a trapped ion and the applied electric field enabled tuning the ring radius $r_0$ to particular values. Vortex rings in superfluids, thanks to the topological robustness due to the quantization condition on the circulation, tend to be very stable (in contrast to their counterparts in hydrodynamics) especially at very low temperatures, where the dissipative effects are very small. On the other hand, the decay of the vortex rings in superfluid $^4$He was used by Bewley and Sreenivasan \cite{BewleySreenivasan} to demonstrate energy dissipation in superfluid $^4$He near the lambda point through the energy transfer from the superfluid to the normal fluid via mutual friction. Direct observation of vortex cores in superfluid $^4$He was accomplished by Bewley et al. \cite{Bewley21} and Fonda et al. \cite{Fonda} by using small solid hydrogen particles as traces in liquid $^4$He.
	
	It may be noted that the generation of vorticity in superfluid $^4$He signifies, on the other hand, the local destruction of superfluidity (Landau \cite{Landau}). So, vortices in superfluid $^4$He essentially behave like classical vortex filaments, barring quantum mechanical features associated with their circulations and extremely thin cores and inclusion of the mutual friction force \footnote{This scenario is however violated in vortex reconnection processes between two neighboring vortex filaments which involve sharp distortions of the vortex filaments (Paoletti el al. \cite{Paoletti}, Bewley et al. \cite{Bewley24}) and the concomitant generation of Kelvin waves associated with helical displacements of the vortex cores (Svistunov \cite{Svistunov}).}. This was very adequately confirmed by the numerical simulations of Schwarz \cite{Schwarz2}, \cite{Schwarz25}.
	
	The extremely thin cores of vortex filaments in superfluid $^4$He lead to a singularity in the vortex self-advection velocity according to the \textit{Biot-Savart} law in hydrodynamics which is resolved by an asymptotic calculation (Da Rios \cite{DaRios}, Arms and Hama \cite{ArmsHama}) called the \textit{local induction approximation} (LIA). Arms and Hama \cite{ArmsHama} used the LIA to investigate the evolution of a perturbed vortex ring in hydrodynamics. Kiknadze and Mamaladze \cite{KM} extended this investigation to consider evolution of a perturbed vortex ring in superfluid $^4$He and found that the mutual friction causes a decay of the perturbation on the vortex ring. The purpose of this paper is to investigate the effect of a toroidal normal fluid flow on a perturbed vortex ring in superfluid $^4$He.\footnote{The Kelvin waves on a linear vortex filament are known to be driven unstable by the normal fluid flow \textit{along} the vortex filament (Shivamoggi \cite{Shivamoggi3}, \cite{Shivamoggi4}).}
	
	\section{Stability of a Vortex Ring in a Superfluid}
	Upon including the mutual friction force (Hall and Vinen \cite{HallVinen7}- \cite{ Vinen9}, Bekarevich and Khalatnikov \cite{BK}) exerted by the normal fluid on a vortex ring, the self-advection velocity of the vortex ring as per the LIA is given by(the HVBK model\footnote{Strictly speaking, the normal fluid flow should be determined as part of the \textit{meso-scale} solution by accounting for the back reaction of the vortices on the normal fluid. However, the HVBK model is valid if the length scales characterizing the flow in question are much larger than the intervortical distance so the vortex lines can be considered to be organized into polarized bundles.}):
	
	\begin{equation}
	\textbf{v}= \gamma \kappa \hat{\textbf{t}}\times \hat{\textbf{n}}+\alpha \hat{\textbf{t}}\times \left( \textbf{U}- \gamma \kappa \hat{\textbf{t}} \times \hat{\textbf{n}} \right)-\alpha' \hat{\textbf{t}}\times\left[\hat{\textbf{t}} \times \left( \textbf{U}- \gamma \kappa \hat{\textbf{t}} \times \hat{\textbf{n}} \right)  \right]
	\label{eq::1}
	\end{equation}
	where \textbf{U} is the normal fluid velocity (taken to be constant in space and time and prescribed (Schwarz \cite{Schwarz2}, \cite{Schwarz25}), $\kappa$ is the average curvature, and $\hat{\textbf{t}}$ and $ \hat{\textbf{n}}$ are unit tangent and unit normal vectors, respectively, to the vortex ring, and $\gamma = \Gamma \ln(c/\kappa a_0)$, where $\Gamma$ is the quantum of circulation, $c$ is a constant of order unity and $a_0 \approx 1.3\times 10^{-8}cm$ is the effective core radius of the filament. $\alpha$ and $\alpha'$ are the mutual friction coefficients which are small (except near the lambda point) so the short-term vortex ring evolution is only weakly affected by the mutual friction. However, it provides for a mechanism to stretch the vortex ring (which is inextensional in the LIA). The mutual friction term assiciated with $\alpha$ plays the dual roles of driving force and drag force (Schwarz \cite{Schwarz2}, \cite{Schwarz25}, Shivamoggi \cite{Shivamoggi3}, \cite{Shivamoggi4}). We drop here the mutual friction term associated with $\alpha'$ because,
	
	\begin{itemize}
		\item $\alpha > \alpha'$ (Vinen and Niemela \cite{VinenNiemela}),
		\item it does not produce physically significant effects in comparison with those produced by the mutual friction term associated with $\alpha$ (Shivamoggi \cite{Shivamoggi3}, \cite{Shivamoggi4}).
	\end{itemize}

	Let us write in cylindrical coordinates (Arms and Hama \cite{ArmsHama}),
	
	\begin{equation}
	\textbf{r} = (r_0+\hat{r})\hat{\textbf{i}}_r+(wt+\hat{z})\hat{\textbf{i}}_z
	\label{eq::2}
	\end{equation}
	where $r_0$ is the unperturbed radius of the vortex ring, $\hat{r}$ and $\hat{z}$ are the deviations from the circular vortex ring in the $r$ and $z$ directions, respectively, and $w$ is the uniform translational self-propelling velocity of the circular ring, given by Kelvin's formula\footnote{According to LIA, an arbitrary vortex filament experiences a self-induced motion, which may be approximated \textit{locally} as that of an osculating vortex ring of radius same as the local radius of curvature of the vortex filament.},
	
	\begin{equation}
	w = \frac{\Gamma}{2 \pi r_0} \ln\left(\frac{8 r_0}{a_0}-\frac{1}{2}\right).
	\label{eq::3}
	\end{equation}
	
	Next, noting (Shivamoggi \cite{Shivamoggi3}, \cite{Shivamoggi4}) that the destabilizing effect of the normal fluid flow is produced by the \textit{toroidal} normal-fluid flow velocity component along the vorticity vector (see Appendix)\footnote{The normal fluid flow along the axis of the vortex ring was found in numerical simulations to keep the vortex ring stable (Kivotides et al. \cite{Kivotides}.}. We therefore take here,
	
	\begin{equation}
	\textbf{U}=U_{\theta}\hat{\textbf{i}}_{\theta}.
	\label{eq::4}
	\end{equation}
	
	Substituting (\ref{eq::2}) and (\ref{eq::4}) in equation (\ref{eq::1}), and neglecting the nonlinear terms, we obtain
	
	\begin{equation}
	\tag{5a}
	\hat{r}_t = \sigma \hat{z}_{\theta \theta}+\alpha \sigma (\hat{r}_{\theta \theta}+\hat{r})-\frac{\alpha}{r_0}U_{\theta}\hat{z}_{\theta}
	\label{eq::5a}
	\end{equation}
	
	\begin{equation}
	\hat{z}_t = -\sigma (\hat{r}_{\theta \theta}+\hat{r})+\alpha \sigma \hat{z}_{\theta \theta}+\frac{\alpha}{r_0} U_{\theta} \hat{r}_{\theta} \tag{5b}
	\label{eq::5b}
	\end{equation}
	where $\sigma \equiv \gamma/r_0^2$.
	\stepcounter{equation}
	
	Looking for solutions of the form,
	
	\begin{equation}
	\hat{q}(\theta, t) \sim e^{i(m \theta -\omega t)}
	\label{eq::6}
	\end{equation}
	equations (\ref{eq::5a}), (\ref{eq::5b}) give,
	
	\begin{equation}
	\omega^2 + i \alpha \sigma \omega (2m^2-1)-\sigma^2 m^2(m^2-1)-i \alpha \sigma \frac{m U_{\theta}}{r_0}(2m^2-1) = 0.
	\label{eq::7}
	\end{equation}
	Noting that $\alpha$ is small, (\ref{eq::7}) gives
	
	\begin{equation}
	\omega \approx - i \alpha \sigma \left[ (2m^2-1) \mp \frac{\frac{U_{\theta}}{r_0}(2m^2-1)}{2 \sigma m (m^2-1)} \right] \pm \sigma m \sqrt{m^2-1}.
	\label{eq::8}
	\end{equation}
	(\ref{eq::8}) shows that the vortex ring develops an instability produced by the \textit{toroidal} normal fluid flow velocity component $U_{\theta}$ (as in the case of a linear vortex filament (Shivamoggi \cite{Shivamoggi3}, \cite{Shivamoggi4}) (see Appendix for a qualitative picture of this instability)). This instability also has qualitative similarities with the \textit{Donnelly-Glaberson instability} (Cheng et al. \cite{Cheng}, Glaberson et al. \cite{Glaberson}) of Kelvin waves on a vortex filament driven by the normal fluid flow \textit{along} the undisturbed vortex filament. If one takes $m=k r_0$ (Kiknadze and Mamaladze \cite{KM}), (\ref{eq::8}) further shows that the growth rate for the present instability is essentially independent of the radius $r_0$ of the vortex ring. This instability would materialize if the time required for this instability to develop is smaller than the time characterizing viscous decay of the toroidal normal fluid flow (which may be taken to be $\sim O(r_0^2/\nu)$, $\nu$ being the kinematic viscosity of the normal fluid), i.e.,
	
	\begin{equation}
	\frac{2 k \nu \left( k^2 r_0^2-1\right)}{\alpha U_{\theta} \left(2 k^2 r_0^2-1\right)}<1
	\label{eq::9}
	\end{equation}
	which favors large toroidal normal fluid flows and large vortex rings.
	
	It may be noted that, if the vortex ring remains closed, $m=1,2,...$, so the motion is periodic around the periphery of the vortex; $m=1$ corresponds to the trivial case of a uniform displacement of a circular vortex ring.
	
	Note that (\ref{eq::8}) reduces,
	
	\begin{itemize}
		\item in the limit $U_{\theta} \Rightarrow 0$, to the result of Kiknadze and Mamaladze \cite{KM} - the effect of mutual friction is now to cause only \textit{decay} of the perturbation on the vortex ring;
		\item in the limit $\alpha \Rightarrow 0$ (the hydrodynamics limit), to the result of Arms and Hama \cite{ArmsHama}.
	\end{itemize}

	\section{Discussion}
	The effect of mutual friction on a self-propelling vortex ring in superfluid $^4$He is to produce a decay of a perturbation imposed on the ring. However, a \textit{toroidal} normal fluid flow is found to drive a vortex ring unstable. This instability has qualitative similarities with the \textit{Donnelly-Glaberson} instability of Kelvin waves on a vortex filament driven by the normal fluid flow \textit{along} the vortex filament. The growth rate of the present instability is further found to be essentially independent of the radius of the vortex ring.

	\section*{Appendix: Stability of a Vortex Filament in a Superfluid}
	\setcounter{equation}{0}
	\renewcommand{\theequation}{A.\arabic{equation}}
	
	Consider a vortex filament aligned essentially along the x-axis in a superfluid (Shivamoggi \cite{Shivamoggi3}, \cite{Shivamoggi4}). Writing in cartesian coordinates,
	
	\begin{equation}
	\textbf{r} = x \hat{\textbf{i}}_x+ \hat{y}(x,t) \hat{\textbf{i}}_y+\hat{z}(x,t) \hat{\textbf{i}}_z
	\label{eq::a1}
	\end{equation}
	taking
	
	\begin{equation}
	\textbf{U} = U_1 \hat{\textbf{i}}_x
	\label{eq::a2}
	\end{equation}
	and neglecting the nonlinear terms, we obtain from equation (\ref{eq::1}),
	
	\begin{equation}
	\tag{A.3a}
	\hat{y}_t = -\sigma \hat{z}_{xx}+ \alpha \sigma \hat{y}_{xx}+\alpha U_1 \hat{z}_x
	\label{eq::a3a}
	\end{equation}
	
	\begin{equation}
	\tag{A.3b}
	\hat{z}_t = \sigma \hat{y}_{xx}+\alpha \sigma \hat{z}_{xx}-\alpha U_1 \hat{y}_x.
	\label{eq::a3b}
	\end{equation}
	\stepcounter{equation}
	
	Putting
	\begin{equation}
	\Phi \equiv \hat{y}+i \hat{z}
	\label{eq::a4}
	\end{equation}
	equations (A.3) give,
	
	\begin{equation}
	i \Phi_t = - \sigma \Phi_{xx}+ \alpha U_1 \Phi_x
	\label{eq::a5}
	\end{equation}
	which may be viewed as a Schr{\"o}dinger type equation for a non-conservative system (Caldirola \cite{Caldirola}, Kanai \cite{Kanai}). If one puts,
	
	\begin{equation}
	\Phi(x,t) = \Psi(x) e^{-i \omega t}
	\label{eq::a6n}
	\end{equation}
	equation (\ref{eq::a5}) leads to
	
	\begin{equation}
	\sigma \Psi_{xx} - \alpha U_1 \Psi_x +\omega \Psi = 0
	\label{eq::a7n}
	\end{equation}
	which represents a harmonic oscillator with \textit{negative} damping.
	
	Looking for solutions of the form
	\begin{equation}
	\Phi(x,t) \sim e^{i(kx-\omega t)}
	\label{eq::a6}
	\end{equation}
	equation (\ref{eq::a5}) leads to:
	
	\begin{equation}
	\omega = i \alpha k U_1+\sigma k^2.
	\label{eq::a7}
	\end{equation}
	(\ref{eq::a7}) shows the destabilization of the circularly-polarized Kelvin waves propagating along the vortex filament\footnote{This result continues to hold in the nonlinear regime as well (Shivamoggi \cite{Shivamoggi3}, \cite{Shivamoggi4}).}.
	
	\section*{Acknowledgments}
	This work was started when the author held a visiting research appointment at the Eindhoven University of Technology supported by a grant from the Burgerscentrum. The author is thankful to Professor Gert Jan van Heijst for his hospitality and helpful discussions. I wish to express my sincere thanks to Professors Carlo Barenghi, Grisha Falkovich, Andrei Golov, Ladik Skrbek, Katepalli Sreenivasan and William Vinen for their valuable remarks. I am also thankful to Dr. Demosthenes Kivotides and Leos Pohl for helpful discussions.

\end{document}